\documentstyle{mn}

\newif\ifAMStwofonts

\ifoldfss
  \ifCUPmtlplainloaded \else
    \NewTextAlphabet{textbfit} {cmbxti10} {}
    \NewTextAlphabet{textbfss} {cmssbx10} {}
    \NewMathAlphabet{mathbfit} {cmbxti10} {} 
    \NewMathAlphabet{mathbfss} {cmssbx10} {} 
  \fi
  \ifAMStwofonts
    \ifCUPmtlplainloaded \else
      \NewSymbolFont{upmath} {eurm10}
      \NewSymbolFont{AMSa} {msam10}
      \NewMathSymbol{\upi}     {0}{upmath}{19}
      \NewMathSymbol{\umu}     {0}{upmath}{16}
      \NewMathSymbol{\upartial}{0}{upmath}{40}
      \NewMathSymbol{\leqslant}{3}{AMSa}{36}
      \NewMathSymbol{\geqslant}{3}{AMSa}{3E}

      \let\leq=\leqslant 
      \let\geq=\geqslant 
    \fi
  \fi
\fi 

\ifnfssone
  \newmathalphabet{\mathit}
  \addtoversion{normal}{\mathit}{cmr}{m}{it}
  \addtoversion{bold}{\mathit}{cmr}{bx}{it}
  \newmathalphabet{\mathbfit} 
  \addtoversion{normal}{\mathbfit}{cmr}{bx}{it}
  \addtoversion{bold}{\mathbfit}{cmr}{bx}{it}
  \newmathalphabet{\mathbfss} 
  \addtoversion{normal}{\mathbfss}{cmss}{bx}{n}
  \addtoversion{bold}{\mathbfss}{cmss}{bx}{n}
  \ifAMStwofonts
    \ifCUPmtlplainloaded \else
      \UseAMStwoboldmath
      \makeatletter
      \new@mathgroup\upmath@group
      \define@mathgroup\mv@normal\upmath@group{eur}{m}{n}
      \define@mathgroup\mv@bold\upmath@group{eur}{b}{n}
      \edef\UPM{\hexnumber\upmath@group}
      \new@mathgroup\amsa@group
      \define@mathgroup\mv@normal\amsa@group{msa}{m}{n}
      \define@mathgroup\mv@bold\amsa@group{msa}{m}{n}
      \edef\AMSa{\hexnumber\amsa@group}
      \makeatother
      \mathchardef\upi=''0\UPM19
      \mathchardef\umu=''0\UPM16
      \mathchardef\upartial=''0\UPM40
      \mathchardef\leqslant=''3\AMSa36
      \mathchardef\geqslant=''3\AMSa3E

      \let\leq=\leqslant 
      \let\geq=\geqslant 
    \fi
  \fi
\fi 

\ifnfsstwo
  \DeclareMathAlphabet{\mathbfit}{OT1}{cmr}{bx}{it}
  \SetMathAlphabet\mathbfit{bold}{OT1}{cmr}{bx}{it}
  \DeclareMathAlphabet{\mathbfss}{OT1}{cmss}{bx}{n}
  \SetMathAlphabet\mathbfss{bold}{OT1}{cmss}{bx}{n}
  \ifAMStwofonts
    \ifCUPmtlplainloaded \else
      \DeclareSymbolFont{UPM}{U}{eur}{m}{n}
      \SetSymbolFont{UPM}{bold}{U}{eur}{b}{n}
      \DeclareSymbolFont{AMSa}{U}{msa}{m}{n}
      \DeclareMathSymbol{\upi}{0}{UPM}{``19}
      \DeclareMathSymbol{\umu}{0}{UPM}{``16}
      \DeclareMathSymbol{\upartial}{0}{UPM}{``40}
      \DeclareMathSymbol{\leqslant}{3}{AMSa}{``36}
      \DeclareMathSymbol{\geqslant}{3}{AMSa}{``3E}

      \let\leq=\leqslant 
      \let\geq=\geqslant 
    \fi
  \fi
\fi 

\ifCUPmtlplainloaded \else
  \ifAMStwofonts \else 
    \def\upi{\pi}
    \def\umu{\mu}
    \def\upartial{\partial}
  \fi
\fi

\title[A $\mu _{0,K}$--$\log h$ relation]
      {A K-band central disk surface-brightness correlation with scale-length 
       for early-type disk galaxies, and the inclination correction}
\author[Alister W. Graham]
       {Alister W.~Graham\thanks{agraham@ll.iac.es}\\
        Instituto de Astrof\'{i}sica de Canarias, La Laguna, E-38200, Tenerife, 
	Spain}
\date{Accepted 1988 December 15.
      Received 1988 December 14;
      in original form 1988 October 11}

\pagerange{\pageref{firstpage}--\pageref{lastpage}}
\pubyear{2001}

\begin{document}
\input{psfig}

\label{firstpage}

\maketitle

\begin{abstract}

The $K$-band light-profiles from two statistically complete, diameter-limited 
samples of disk galaxies have been simultaneously modelled with a 
seeing-convolved S\'{e}rsic $r^{1/n}$ bulge and a seeing-convolved exponential disk. 
This has enabled an accurate separation of the bulge and disk light, 
and hence an estimate of the central disk surface brightness $\mu_{0,K}$ 
and the disk scale-length $h$. 
There exists a bright envelope of galaxy disks in the $\mu_{0,K}$--$\log h$ 
diagram; for the early-type ($\leq$Sbc--Sc) disk galaxies $\mu_{0,K}$ is 
shown to increase with $\log h$, with a slope of $\sim$2 and a correlation 
coefficient equal to 0.75.   This relation exists over a range of disk 
scale-lengths from 0.5 to 10 kpc ($H_{\rm 0}$$=$75 km s$^{-1}$ Mpc$^{-1}$).  
In general, galaxy types Scd or later are observed to deviate from this relation; 
they have fainter surface brightnesses for a given scale-length. 

With a sub-sample of 59 low-inclination ($i$$\leq$50$^{\rm o}$) and 29 
high-inclination ($i$$\geq$50$^{\rm o}$) galaxies having morphological types 
ranging from S0 to Sc, the need for an inclination correction to the $K$-band 
disk surface brightness is demonstrated. 
Certain selection criteria biases which have troubled previous surface brightness
inclination tests (for example, whether the galaxies are selected from a magnitude- 
or diameter-limited sample) do not operate in the $\mu_{0,K}$--$\log h$ diagram. 
Measured central disk surface brightnesses are found to be 
significantly ($>$5$\sigma$) brighter for the high-inclination disk galaxies than 
for the low-inclination disk galaxies.  With no surface brightness inclination 
correction or allowance 
for the trend between $\mu_{0,K}$ and $\log h$, the standard deviation to the 
distribution of $\mu_{0,K}$ values is $\sim$1 mag arcsec$^{-2}$, while the standard 
deviation about the mean
$\mu_{0,K}$--$\log h$ relation decreases from 0.69 mag arcsec$^{-2}$, when no 
inclination correction is applied, to 0.47 mag arcsec$^{-2}$ when the inclination 
correction is applied.  
Possible changes to the disk scale-length with inclination, due to radial 
gradients in the disk opacity, have been explored.  The maximum possible size of 
such corrections are too small to provide a valid explanation for the 
difference between the low- and high-inclination disk galaxies in the 
$\mu_{0,K}$--$\log h$ diagram. 

\end{abstract}


\section{Introduction}

Over the last three decades, the validity of Freeman's law -- 
the observation that the $B$-band central disk surface brightness of spiral
galaxies is roughly constant (Freeman 1970) -- has been investigated many 
times.  The interest generated by this work 
is not surprising given its implications for the formation and nature
of disk galaxies.  For instance, if the central $M/L_B$ ratio is 
constant in spiral galaxies, it would imply a constant central mass surface 
density which would provide a strong constraint to any potential disk galaxy 
formation mechanism.  
Subsequent studies observed the scatter in Freeman's law to increase, 
notably from the inclusion of fainter late-type ($>$Sc) spiral galaxies 
(van der Kruit 1987), vindicating the argument by Disney (1976, 1980) that Freeman's
law was an artifact of selection effects.  This result was expanded upon 
by de Jong (1996b), who argued that Freeman's law must be redefined as 
de Jong found only an upper-limit to the central disk surface brightness term.  
Similarly, studies of low surface
brightness galaxies reveal a class of spiral galaxies with central disk surface 
brightnesses lower than the canonical Freeman value (Longmore et al.\ 
1982; Bothun et al.\ 1987; Davies et al.\ 1988; van der Hulst et al.\ 1993; 
McGaugh \& Bothun 1994; de Blok, van der Hulst \& Bothun 1995; de Blok, 
van der Hulst \& McGaugh 1996; McGaugh 1996; Bothun, Impey \& McGaugh 1997) 
and with no indication of a lower-limit.  

An important consideration when measuring the central surface brightness of a 
disk is the question of the disk's opacity.  
Extinction due to dust is an order of magnitude less in the $K$-band 
than the $B$-band.  The $K$-band also has the 
advantage of tracing the old stellar population, and not the few percent
(by mass) of young stars which dominate the light in the $B$-band. 
However, at least in the $H$-band, dust can still have a measurable effect 
on the disk surface brightness distribution (Aoki et al.\ 1991; 
Barnaby \& Thronson 1992; Moriondo, Giovanelli \& Haynes 1998). 
Amongst other things, this paper attempts to answer the question:\  
What inclination correction, if any, needs to be applied to
the observed $K$-band surface brightnesses (and magnitudes) of disk galaxies?
Applying the right inclination correction is important as it influences 
statistical studies of the properties of spiral galaxies. 
It affects derivations of the luminosity function (Leroy \& Portilla 
1998), and hence galaxy magnitude and 
redshift number counts, and it also affects corrections to the Tully-Fisher 
relation for studies of peculiar velocity flows and large-scale structure. 

One of the difficulties in using a galaxies central surface brightness is 
that it is a combination of both disk and bulge light.  For a galaxy sample 
that includes Sa through to Sd or Sm galaxies, the contribution of the central 
bulge light to the central disk light can span a large range of values. 
Even for the Sc galaxies the bulge can contribute almost no extra light at the 
galaxy centre or it can be several magnitudes brighter than the disk. 
Kormendy (1977), Phillipps \& Disney (1983), and later Davies (1990) argued 
that failure to accurately model the contribution of
bulge light may place constraints on the true range
of central disk surface brightness values which are measured. 
Although the physical applicability of the parameter space explored in these 
authors models has been questioned (de Jong 1996b), at some level, the issue 
of bulge contamination remains.  

\begin{figure}
\centerline{\psfig{figure=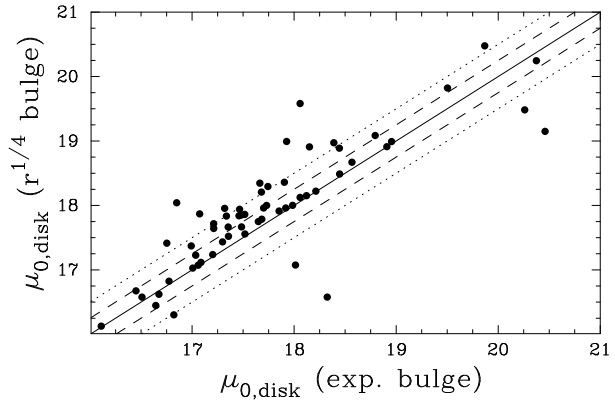,width=8.0cm,angle=-90}}
\caption{The central $K$-band disk surface brightness, $\mu _{\rm 0,disk}$, 
derived by 
simultaneously fitting an $r^{1/4}$ bulge with an exponential disk, and an 
exponential bulge with an exponential disk are plotted against each other
(fitting and data from de Jong 1996a).  The dashed and dotted lines mark 
the 0.25 and 0.50 mag arcsec$^{-2}$ deviations from agreement.}
\label{fig0}
\end{figure}

Many of the pioneering papers which fitted both a bulge and disk generally modelled the 
bulge with a de Vaucouleurs (1948) $r^{1/4}$ profile (Kormendy 1977; Burstein 1979;
Boroson 1981, 1983; Kent 1985; Schombert \& Bothun 1987).  The risks of this 
assumption were mentioned by these authors and departures noted.  de Jong (1996a) 
actually rejected the $r^{1/4}$ bulge profile in favour of an exponential
bulge profile, which, in general, gave a statistically better fit to his sample 
of 86 disk galaxies.  Today, however, the range of bulge profile shapes
has been investigated (Andredakis, Peletier \& Balcells 1995, hereafter APB95; 
Moriondo, Giovanardi \& Hunt 1998), and application of the classical 
exponential profile or $r^{1/4}$ law to a bulge which is better described 
by some other light profile shape is known to introduce errors into the 
structural parameters of the galaxies (Graham \& Prieto 1999, Graham 2001). 
While the effect is most dramatic on the bulge parameters, 
differences to the estimated central disk surface brightness can vary by 0.5 
mag arcsec$^{-2}$ or more, depending on the assumed bulge profile shape.  
To illustrate this, Figure~\ref{fig0} shows the $K$-band central disk surface 
brightness estimates from de Jong (1996a), obtained when modelling the bulge with 
first an $r^{1/4}$ profile and then an exponential profile.   One can see that 
considerable differences arise when forcing a restrictive bulge model that has 
no physical basis.  In an effort to separate the bulge light from the disk 
profile as accurately as possible, in this analysis a seeing-convolved $r^{1/n}$ 
bulge profile has been simultaneously fitted with a seeing convolved exponential 
disk profile.

With a sample of low- and high-inclination galaxies, the selection of which is 
presented in Section 2, this paper identifies a considerable source of scatter 
to the $K$-band central disk surface brightness -- namely, the dependence on the 
disk scale-length  (preliminary results have already been presented in Graham 
2000a,b).  Section 3 provides a description of the standard inclination 
correction while Section 4 attempts to provide a quantitative measure to the size of the 
$K$-band surface brightness inclination correction needed for spiral galaxy disks. 
In Section 5 the consequences of an opacity dependence on radius, such that disks
are more optically thick towards the centre, is explored and found not to be important 
in the $\mu_{0,K}$--$\log h$ diagram.  Section 6 provides a discussion of 
the results and Section 7 presents the main conclusions from this investigation.

\section{Data}

The data used in this investigation are from two statistically complete, diameter-limited 
samples of galaxies selected from the Uppsala General Catalogue of Galaxies (UGC, Nilson 
1973).  A selection of galaxies with inclinations less than 50$^{\rm o}$ were 
investigated by de Jong \& van der Kruit (1994) while a somewhat complimentary 
sample of galaxies with inclinations greater than 50$^{\rm o}$ were investigated by 
Balcells \& Peletier (1994) and later APB95.  Both studies avoided disturbed or 
interacting systems. 
$K$-band images taken with UKIRT's IRCAM2, resulted in a volume-limited
sample of 86 low-inclination spiral galaxies; for which the data reduction 
is described in de Jong \& van der Kruit (1994).\footnote{The surface brightness
profiles for the low-inclination sample, obtained from the average surface brightness 
on ellipses of constant ellipticity and position angle, can be found at
http://cdsweb.u-strasbg.fr/htbin/Cat?J/A+AS/118/557.}
Images taken with UKIRT's IRCAM3 resulted in a sample of 30 early-type ($T$$<$5) 
high-inclination spiral galaxies, for which the data reduction procedures 
can be found in APB95 and  Peletier \& Balcells (1997).\footnote{The surface
brightness profiles for the high-inclination sample can be found at 
http://www.elsevier.nl/gej-ng/10/33/29/22/23/23/article.html.
(The K-band light-profiles were taken from the ellipse fits, not the wedge profiles.)}
Due to the complimentary inclination selection-criteria, no common galaxies have 
been observed -- preventing a check of the photometric zero-point.  This is
unfortunate, however, although calibration offsets of 0.05 mag are not uncommon 
between different data sets, this would not significantly impact the results of 
this investigation and is certainly tolerable. 

There is a difference between the diameter-limited selection criteria of each sample. 
The low-inclination galaxies were selected to have a red major-axis diameter 
$\geq$2.0 arcmin, while the high-inclination galaxies were selected to have a blue 
major-axis diameter $>$2.0 arcmin.  The difference to the final high-inclination sample, 
had the selection criteria for the larger low-inclination sample been used, is small. 
Only one galaxy (UGC 9016) with a blue major-axis diameter greater than 2.0 arcmin 
and having a red major-axis diameter less than 2.0 arcmin was included in the final 
sample of 30 galaxies by APB95 -- this galaxy is easily excluded from this analysis.  
Two galaxies (UGC8485 and UGC 10713) with a red major-axis diameter $\geq$2.0 arcmin
but with a blue-major-axis $\leq$2.0 arcmin would have been included if the diameter 
selection criteria was the same.  

APB95 applied further selection criteria which de Jong \& van der Kruit (1994)
did not, but which can be applied here to make the samples as similar as possible.  
While de Jong \& van der Kruit (1994) selected disk galaxy types from S0 to Irr,
APB95 restricted their sample to galaxy types earlier than Sc. 
APB95 also applied a magnitude selection criteria, such that only galaxies with
$B_T$$<$14.0 mag were included, and they excluded galaxies with bars. 
These same criteria are subsequently applied individually and all at once to the 
low-inclination sample to see if this affects the results of this investigation.  
The only major concern is that APB95 excluded 3 galaxies solely because they
were observed to be very dusty, in the optical, all the way to the centre.  
This may well have an affect on the mean surface brightness of the high inclination
sample.  However, this still left a sample of 30 high-inclination galaxies and so it's 
impact is not expected to be terribly substantial but should, nonetheless, be noted.

Using the same algorithm, all galaxies have been simultaneously modelled with 
a seeing convolved S\'{e}rsic (1968) bulge profile and a seeing convolved 
exponential disk.  The model parameters obtained are the bulge effective 
half-light radius ($r_e$), the bulge surface brightness at this radius 
($\mu_e$), the shape of the bulge profile ($n$), the central disk surface 
brightness ($\mu_0$), and the disk scale-length ($h$).  The disk scale-length
is that appropriate to the major-axis, and is not the geometric mean 
between the major- and minor-axis -- a quantity shortened through inclination. 
A presentation of the fits to the low-inclination galaxies can be found in Graham 
(2000).  Three of these galaxies have optical but no $K$-band data, and 5 
were imaged under non-photometric conditions.  Of the 86 low-inclination disks, 65 
have morphological type $T$$\leq$5 (42 have $T$$\leq$4) -- taken from the RC3 
(de Vaucouleurs et al.\ 1991).  Only two galaxies having $T$$\leq$5 could not 
be reliably modelled.  These are UGC 8279 which has no apparent bulge -- causing the 
bulge/disk decomposition to fail -- and UGC 6028 whose disk has two distinct slopes.  
This left a sample of 59 low-inclination disk galaxies with morphological type 
$T$$\leq$5 (39 with $T$$\leq$4).  For the sample of 59 galaxies, a $V/V_{\rm max}$ 
test gave an average value of
0.53 -- in agreement with the expected value of 0.5$\pm$$1/\sqrt{12\times N}$ 
for a statistically complete, diameter-limited sample ($=$0.5$\pm$0.04 when $N$$=$59).
Similarly, the smaller sample of 39 galaxies represents a statistically complete, 
volume-limited sample of $T$$\leq$4 disk galaxies, with the average 
$V/V_{\rm max}$$=$0.49.  Although the remaining later type spiral galaxies 
are included in Figure~\ref{fig2} -- to help highlight that the selection criteria 
are not clearly responsible for the observed trends -- these galaxies have not been 
used in the statistical analysis which follows.  
All high-inclination galaxies, excluding UGC 9016, have been included. 
For these, the $V/V_{\rm max}$ test gave an average value of 0.51 -- within the range 
0.5$\pm$0.05 expected for a sample of 29 galaxies that is randomly distributed in space. 

The central surface brightnesses have been uniformly corrected 
for a) Galactic extinction using the composite IRAS and COBE/DIRBE dust 
extinction maps of Schlegel, Finkbeiner \& Davis (1998) -- data obtained 
from NED b) $(1+z)^4$ cosmological redshift dimming and c) K-correction 
using heliocentric velocities (Poggianti 1997).   For the low-inclination galaxies, 
the average size of these corrections was relatively small at -0.02, -0.05, 
and 0.02 mag respectively, with maximum corrections of -0.06, -0.12, and 0.04 mag.
For the high-inclination sample, these corrections were again small, with average values 
of -0.02, -0.03, and 0.01, and they do not significantly alter the results of this paper.

In this paper a value of $H_{\rm 0}$$=$75 km s$^{-1}$ Mpc$^{-1}$ has been used. 
For both the low- and high-inclination galaxy samples, their distances were
derived using the heliocentric velocities from NED which were then corrected 
for Virgo-infall using the 220 model of Kraan-Korteweg (1986).

\section{The standard inclination correction}

In the case of a transparent disk, stellar light is radiated equally
in all directions. Consequently, the viewing angle (inclination) will 
not effect the observed magnitude of the galaxy.  However, transparent 
disks that are seen at higher inclinations (with inclination $i$ 
increasing from face-on to edge-on) 
project to a smaller area on the sky and consequently their surface brightness 
(intensity/area) will appear brighter than if they had been viewed face-on.   
In contrast to this, for an optically thick disk the observed surface 
brightness will be independent of viewing angle, as the line-of-sight 
depth into the disk will be the same for all inclinations and the observed 
intensity per unit area will remain fixed. 
For optically thick disks viewed at higher inclinations, the total galaxy 
magnitude decreases as the projected area of the galaxy is decreased
from its face-on value. 

This situation has been expressed quantitatively with a simple model such that
\begin{equation}
\mu_{{\rm inc}} = \mu_{{\rm face-on}} + 2.5 C \log (b/a) \label{eqn1}
\end{equation}
where $b/a$ is the ratio of the semi-minor to semi-major axis of the disk. 
For an optically thick disk $C$$=$0, and for an optically thin disk 
$C$$=$1.  Of course, spiral disks may well be described by some intermediate
degree of opacity, where $C$ would take on a value somewhere between 
0 and 1.  The value of $C$ is also known to be a function of wavelength 
since its origin is dust. 

One can envisage a number of special locations for dust within a disk that
would complicate the opacity test.  
In the text that follows, I shall refer to a disk that behaves transparently
in an inclination test as a `transparent disk'.  The disk may in fact not be 
transparent, or may be only semi-transparent.  In one case of interest, 
dust may reside in the disk with a scale-height less than that of the stars.  
If so, as noted by Disney et al.\ (1989), the outer parts of the disk (in the 
$z$-direction) may be transparent while sandwiched between these two outer layers 
is obscuring dust.   As the inclination increases, the outer layers, as opposed to 
the disk as a whole, will cause the observed surface brightness to brighten above 
its face-on value (Davies et al.\ 1993; Davies, Jones \& Trewhella 1995).  
Another case of interest is if a dust ring was to encircle the disk.  Although 
unlikely, this may occur to a degree by dust in the outer spiral arms (Huizinga 1994). 
For face-on galaxies, and galaxies at intermediate inclinations, this would simply 
introduce a wiggle into the outer light profile, but for those galaxies which are 
very edge-on this would create a biased impression of the global opacity of the disk.  

Irrespective of the location or distribution of the dust, all future 
reference to the value of $C$ is made only in regard to the required, 
statistically averaged, inclination correction, and not to the nature 
of the disk as a whole.

\section{Application of the correction}

\subsection{A constant central disk surface brightness -- Freeman's law}

The inclination correction was uniquely determined for each galaxy depending on 
the ellipticity of its disk.  
With $C$$=$0, the inclination correction to the surface brightness 
term is zero.  With $C$$=$1, the average size of the inclination correction 
for the low-inclination sample of spiral galaxies is 0.26 mag arcsec$^{-2}$, and 
the maximum correction 0.58 mag arcsec$^{-2}$.  
For the high-inclination sample of galaxies, when $C$$=$1, the average 
correction is 1.34 mag arcsec$^{-2}$ and the maximum correction is 1.92 mag 
arcsec$^{-2}$ for a galaxy with a disk ellipticity ($\epsilon$$=$1-$b/a$) equal 
to 0.83.
In passing, it is noted that if some of the galaxy 
disks do not posses circular symmetry but are intrinsically elongated, then this 
will introduce a source of scatter into the inclination corrections.  One may 
therefore be able to further reduce the scatter in the central disk surface 
brightness distribution by using kinematic, rather than photometric, inclinations 
(Andersen et al.\ 2000).  Although, these differences are greatest at small 
inclinations where the size of the inclination corrections are small. 

\begin{figure}
\centerline{\psfig{figure=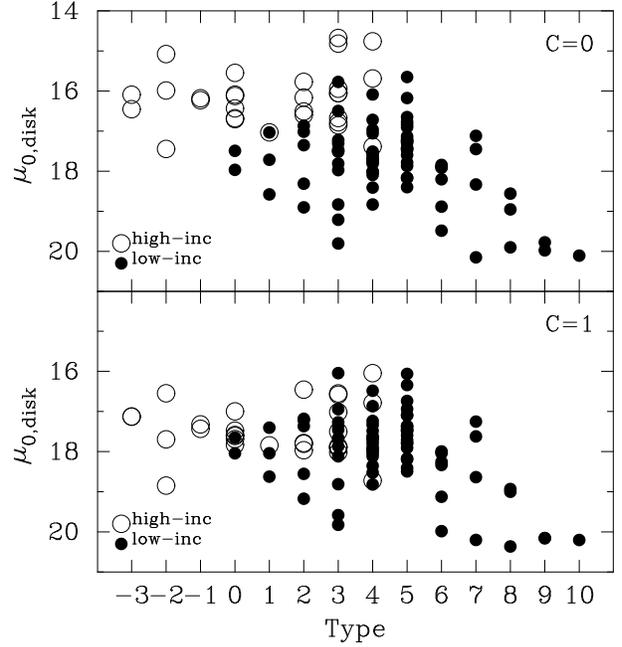,width=8.0cm,angle=0}}
\caption{The central $K$-band disk surface brightness is plotted against galaxy 
morphological Type, with no inclination correction ($C$$=$0) and the maximum 
inclination correction ($C$$=$1) applied. (The value of $C$ comes from the
inclination correction $2.5C\log(b/a)$ given in the text.)
The high-inclination ($i$$>$50$^{\rm o}$) disk galaxy sample are represented with
the open circles, and the low-inclination ($i$$<$50$^{\rm o}$) disk galaxy sample
are represented by the filled circles.}
\label{fig1}
\end{figure}

The scatter in the distribution of central disk surface brightnesses 
is known to be larger in the $K$-band than the B-band, and $\mu_0$ is
known to increase for galaxy types $T$$>$5-6 (Grosb\o l 1985; Valentijn 1990; 
Peletier et al.\ 1994; de Jong 1996b; Tully \& Verheijen 1997; de Grijs 1998)
and this is also seen with the current sample.   In Figure~\ref{fig1}, 
when $C$$=$1, $\mu_{0,K}$ displays no trend with morphological 
type $T$$\leq$5.  Given the similarity of central disk surface brightness 
between the $T$$=$5 galaxies and the earlier types, they have been 
included in the statistical analysis which follows so as to increase galaxy numbers
and improve the statistics. 
In later comparison studies between the low- and high-inclination samples, the 
affect of removing the $T$$=$5 galaxies, which are not present amongst the 
high-inclination sample, is explored. 

For the low-inclination galaxies with type $T$$\leq$5, the mean
value of $\mu_{0,K}$ is 17.45$\pm$0.76 ($C$$=$0) and 17.72$\pm$0.74 ($C$$=$1)
For the high-inclination sample, the mean value of $\mu_{0,K}$ is 16.10$\pm$0.70  
($C$$=$0) and 17.45$\pm$0.65 ($C$$=$1).
%
%
For the combined sample of galaxies with $T$$\leq$5, the mean central disk 
surface brightness is 17.01$\pm$0.98 ($C$$=$0) and 17.63$\pm$0.72 ($C$$=$1). 
The inclusion of galaxy types with $T$$>$5 increases both the mean value of 
$\mu_{0,K}$ and the associated error.

\subsection{A central disk surface brightness correlation with scale-length}

\begin{figure}
\centerline{\psfig{figure=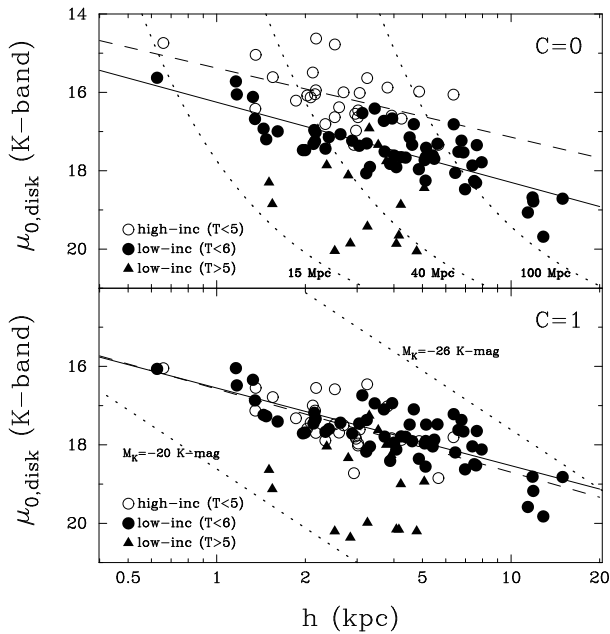,width=8.0cm,angle=0}}
\caption{The central $K$-band disk surface brightness is plotted against
the scale-length of the disk for the case when $C$$=$0 (i.e.\ no
inclination correction) and when $C$$=$1.
The solid line is the linear regression of $\mu_{0,K}$ on $\log h$ for the
low-inclination galaxies with type $T$$\leq$5.  The dashed line is the regression
for the high-inclination galaxy sample.  The dotted curves in the upper panel
indicate the 2.0 arcmin diameter (1.0$\arcmin$ radius) selection limit at the
24.7 $R$-mag arcsec$^{-2}$ isophote for galaxies closer than 15, 40 and 100 Mpc,
and assuming galaxies have perfect exponential disks with $R$-$K$=2.25.
The dotted lines in the lower panel mark where disks of a fixed luminosity
reside.  A Hubble constant of 75 km s$^{-2}$ Mpc$^{-2}$ has been used.}
\label{fig2}
\end{figure}

A plot of $\mu_{0,K}$ against $\log h$, where $h$ is the scale-length of the disk, 
reveals -- for the early-type disk galaxies -- a correlation between these 
two parameters (Figure~\ref{fig2}).  One can also see, when no inclination 
correction is applied ($C$$=$0), that for a given size (scale-length) of a disk, 
the high-inclination sample are observed to have brighter values of $\mu_{0,K}$ than 
the low-inclination sample -- indicating the transparent behaviour of at least
the early-type disks.  The decision to include the $T$$=$5 galaxies in the 
low-inclination sample does not appear to affect the observed 
$\mu_{0,K}$--$\log h$ relation, where-as the later galaxy types do show a 
departure from the relation defined by the earlier-type galaxies.  A 
discussion of the influence of the selection criteria on this apparent 
correlation is deferred to Section 6. 

After application of the $C$$=$1 inclination correction, a Pearson correlation 
coefficient of 0.75 exists between $\mu_{0,K}$ and $\log h$ (kpc) for the combined 
sample of low- and high-inclination early-type ($T$$\leq$5) disk galaxies. 
The previous standard deviations in the distribution of $\mu_{0,K}$ are 42\% 
and 53\% higher than the standard deviations about the mean $\mu_{0,K}$--$\log h$ 
relation -- which are 0.69 mag arcsec$^{-2}$ (when $C$$=$0) and 0.47 mag 
arcsec$^{-2}$ (when $C$$=$1).

From Figure~\ref{fig2}, it is clear that when no correction is made for
inclination ($C$=0) the low- and high-inclination data sets have 
different mean central surface brightnesses.
Subtracting the mean $\mu_{0,K}$--$\log h$ relation, one can remove the dependence
of surface brightness on galaxy size before computing the difference in mean
values of $\mu_{0,K}$ for the two samples.  A Student's $t$-test on the 
two residual data sets (low-inclination and high-inclination) reveals 
significantly different means.  Using the F-test for 
the similarity of variances, these residual data sets have different variances 
at the 1.5$\sigma$ significance level.  
Student's $t$-test was therefore applied twice: assuming equal and then  
unequal variances.   The probability that the value of 
$t$ from Student's test could be as large as it is just by chance is smaller 
than $10^{-7}$ -- indicating that the difference in means is highly significant.  
In addition to subtracting the mean $\mu_{0,K}$--$\log h$ 
relation for the combined sample, the mean $\mu_{0,K}$--$\log h$ relation for the 
low-inclination sample, and also for the high-inclination sample, were separately 
subtracted and Student's $t$-test applied again. 
From all six $t$-test's (equal and unequal variances, and using the three
different mean $\mu_{0,K}$--$\log h$ relations) the probability that the 
difference in means could be as large as it is remained less than $10^{-7}$. 

With $C$$=$1, the same tests were again applied.  This time, the variances 
of the two residual data sets were not only found to be equal at the 1$\sigma$ 
level, but were notably reduced for the high-inclination sample.  The mean 
values of $\mu_{0,K}$ 
from the residual data sets -- having subtracted the mean $\mu_{0,K}$--$\log h$ 
relation -- were found to be consistent with each other at well within
the 1$\sigma$ confidence limit. That is, application of Student's $t$-test
reveals that with $C$$=$1, the mean surface brightness offset between the 
low- and high-inclination galaxies in the $\mu_{0,K}$--$\log h$ plane is 
not significant. 

To compute the mean offset, a modified application of the Working-Hotelling (1929) 
confidence bands was used (Feigelson \& Babu 1992). 
This method of analysis performs a linear regression on one data set 
(the calibration set) and then, given the $x$-values ($\log h$) of the
second data set, predicts their $y$-values (in our case $\mu_{0,K}$) and their 
associated uncertainty.  The difference between these predicted values 
and the 
observed values is a measure of the offset.  
The standard deviation of this distribution of differences divided 
by $\sqrt N$ gives the uncertainty on the mean of the offset.  
With $C$$=$0 or $C$$=$1 the low- and high-inclination data sets are 
consistent (at the 1$\sigma$ level) with being parallel to each other -- 
a requirement of the Working-Hotelling technique.  
When $C$$=$0 the mean vertical offset is -0.97$\pm$0.11 mag arcsec$^{-2}$, and 
when $C$$=$1 the mean offset is +0.10$\pm$0.10 mag arcsec$^{-2}$.\footnote{The 
Working-Hotelling confidence bands allow one to obtain two estimates of 
the offset by using in turn both data sets as the `calibration' data set.  
The mean offset was not significantly different when this was done.}

The vertical scatter about the $\mu _{0,K}$--$\log h$ relation for the high-inclination 
sample is reduced by 20\% when the $C$$=$1 inclination correction is applied.  
Further support for the validity of the inclination correction comes from 
the improved agreement in slope of the $\mu _{0,K}$--$\log h$ relation between 
the low- and high-inclination data sets.  For the low-inclination spirals,
an ordinary least-squares regression of $\mu _{0,K}$ on $\log h$ reveals a slope of 
2.05$\pm$0.23 ($C$=0) and 1.98$\pm$0.22 ($C$=1). 
For the high-inclination spirals, the slope is 1.76$\pm$0.59 ($C$=0) and 
2.12$\pm$0.41 ($C$=1). 
The error estimates for the slopes are slightly larger 
($\sim$5-10\%) than those obtained from the standard analytical solution 
which, for small sample sizes ($N$$\leq50$), can under-estimate the size 
of the error.  The errors shown here are the largest standard deviations 
obtained from jackknife and bootstrap resampling of the data.  
For the combined data sets, with $C$$=$1, the mean slope is 1.96$\pm$0.18
with a scatter about the $\mu _{0,K}$--$\log h$ relation of 0.47 mag arcsec$^{-2}$.

With $C$$=$0.91, the mean offset between the two samples was 0.00, but this 
did not result in a significantly different slope or scatter to the 
$\mu_{0,K}$--$\log h$ relation obtained when $C$$=$1.  For the low-inclination 
sample, with $C$$=$0.91, the slope was 1.99$\pm$0.21 with a scatter of 0.47 mag 
arcsec$^{-2}$.  For the high-inclination sample the slope was 2.08$\pm$0.48 with 
a scatter of 0.49 mag arcsec$^{-2}$, and for the combined sample the slope was 
2.01$\pm$0.18 with scatter 0.47 mag arcsec$^{-2}$ and a Pearson's correlation
coefficient of 0.76. 
Completely independent of the two inclination bins ($i$$<$50$^{\rm o}$ or 
$i$$>$50$^{\rm o}$), the scatter about the $\mu_{0,K}$--$\log h$ relation 
reduced to a minimum as the value of $C$ increased to 0.9-1.0. 

If one wishes to use the $K$-band $\mu_{0}$--$\log h$ relation for distance estimates,
then one should perform a linear regression which minimizes the distance 
dependent quantity $\log h$ (kpc) against the distance independent quantity 
$\mu_{0,K}$.\footnote{Strictly speaking, the surface brightness term is dependent on 
the redshift via the K-correction and redshift dimming, but these corrections, or 
more importantly, realistic errors in these corrections due to errors in redshift
are less than 0.01 mag arcsec$^{-2}$.} 
Doing this, with $C$$=$1, the resulting standard deviation is a somewhat large 
0.18 dex in $\log h$ which corresponds to a mean distance uncertainty of $\sim$50\%.  

\subsubsection{Sub-sampling the data}

The above analysis was repeated several times, culling each data set to make them
more similar to each other -- and testing if these steps caused any significant
changes (at the 1$\sigma$ level).  Firstly, 
the Sc galaxies included in the low-inclination sample were removed. This had 
no significant effect on any of the above numbers.  Next, because the low-inclination
sample has only two S0 galaxies (both with type $T$$=$0) while the high-inclination
sample includes S0 galaxy types from $T$$=$-3 to 0, all S0 galaxies and Sc galaxies
were removed from both samples, leaving only types $T$=1 to 4.  Again, the 
numbers obtained were not significantly different with those obtained using the 
full sample.  With $C$$=$1 the slope of the $\mu _{0,K}$--$\log h$ relation changed 
to 2.03$\pm$0.24 with a slightly greater Pearson correlation coefficient of 0.77. 

The high-inclination galaxy sample of APB95 excluded galaxies with a prominent bar --
of the 29 galaxies used from this sample, NED still lists 8 of these with having a bar. 
Therefore, the above tests were repeated 
excluding those galaxies from the 59 low-inclination galaxies which de Jong (1996a)
considered to have a prominent bar -- which he modelled with an additional bar 
component.  None of the results differed at more than the 1$\sigma $ significance 
level and usually well under this.  
Excluding the prominently barred galaxies, the mean vertical offset 
between the $\mu_{0,K}$--$\log h$ relation 
for the low- and high-inclination samples is -1.00$\pm$0.11 ($C$$=$0) 
and +0.07$\pm$0.10 ($C$$=$1).   The mean slope of the $\mu_{0,K}$--$\log h$ 
relation for the combined data sets changed to 1.83$\pm$0.21 ($C$$=$1) with a scatter 
of 0.49 mag arcsec$^{-2}$. 
Next, the 8 galaxies from the high-inclination sample listed as having bars were 
removed as was every galaxy in the low-inclination sample which was listed in NED 
as having a bar.  Once again, the above results did not significantly differ, and 
it can be concluded that these results are not biased by the presence of a bar in 
some of the sample members.  

APB95 further restricted their sample to galaxies listed in the UGC with $B_T$$<$14.0 mag.
This same restriction was applied here to the low-inclination sample to test if this
may be influencing the results -- it does not.  
Lastly, the above criteria were all applied at once to see if they may be working 
together to bias the above results.  All S0 and Sc galaxies were excluded, as
were all galaxies listed in the UGC with $B_T$$<$14.0 mag and all galaxies 
considered to have a prominent bar.  
Ignoring the dependence of $\mu_{0,K}$ on $\log h$, the mean value of $\mu_{0,K}$ for
the low-inclination sample was 17.43$\pm$0.71 ($C$$=$0) and 17.67$\pm$0.66 ($C$$=$1).
For the high-inclination sample these values were 16.05$\pm$0.82 ($C$$=$0) and 
17.42$\pm$0.75 ($C$$=$1).  This heavy sample truncation had the greatest effect on 
the $\mu_{0,K}$--$\log h$ relation, whose slope became 1.80$\pm$0.34 ($C$$=$1) -- 
still, however,
consistent with the solution obtained with the larger galaxy sample.  
The scatter about the $\mu_{0,K}$--$\log h$ relation changed to 
0.87 mag arcsec$^{-2}$ ($C$$=$0) and 0.51 mag arcsec$^{-2}$ ($C$$=$1).

To continue in this vain, and address constructive criticism, 
next, those galaxies at the tails of the scale-length distribution were excluded.
Removing the one galaxy with a disk scale-length less than 1 kpc from the 
high-inclination sample changed the slope of the $\mu_{0,K}$--$\log h$ relation
for the high-inclination sample to 2.00$\pm$0.59 ($C$$=$1).  Removing the 1 
galaxy from the
low-inclination sample with a disk scale-length less than 1 kpc had practically no 
affect at all.  At the other end of the distribution, 
five galaxies from the low-inclination sample have scale-lengths larger than 10 kpc
($H_{\rm 0}$$=$75 km s$^{-1}$ Mpc$^{-1}$).  Although, when using the same Hubble 
constant as de Jong  ($H_{\rm 0}$$=$100 km s$^{-1}$ Mpc$^{-1}$), only 2 galaxies have a
disk scale-length greater than 9 kpc and only 1 galaxy has a disk scale-length 
greater than 10 kpc.  Still, these five galaxies were excluded to test if they 
may be biasing the $\mu_{0,K}$--$\log h$ correlation.  Doing this, the slope 
of the combined data set was reduced to 1.67$\pm$0.21 ($C$$=$1).  Excluding 
these five galaxies, and 
those two galaxies with disk scale-lengths less than 1 kpc, the slope of the
combined data set became 1.55$\pm$0.23 ($C$$=$1), with a Pearson correlation 
coefficient of 0.60.  This is still a highly significant result.  The Spearman 
rank-order correlation coefficient is 0.58 -- with a significance of 
2.10$^{-8}$.  Similarly, the sum-squared difference of ranks deviates from the 
expectation value in the null hypothesis of uncorrelated data sets by more than 
5 standard deviations. 

As a final check, the high-inclination sample had previously been independently 
modelled by Khosroshahi et al.\ (2000) using a two-dimensional image decomposition 
technique that fitted a seeing convolved S\'{e}rsic bulge and exponential disk.  
Applying the inclination correction to the data in their Table 1, and the 
various other surface brightness and Virgo-infall corrections mentioned earlier, 
the above analysis was repeated again and found to be fully consistent with the 
results presented above.

\section{Possible scale-length changes with inclination}

For a galaxy with a globally uniform disk opacity, the surface brightness profile 
will brighten with inclination (or remain the same if it is optically thick) by 
the same amount at all radii.  The scale-length of the disk would therefore be 
independent of inclination.  If, however, the opacity of the disk changes 
with radius, then the scale-length could change as the galaxy was viewed at different 
inclination angles.
This scenario opens the door to the possibility that the inclined galaxies in
Figure~\ref{fig2} may not have simply moved along lines of constant scale-length.
In the extreme, one may postulate that disks are optically thick at their
centres and $\mu_{0,K}$ does not change with inclination but instead the scale-length
changes to produce the shift in Figure~\ref{fig2}.  However, rather than acting 
as a hindrance to the question of disk opacity, this effect is rather fortuitous 
as it actually enables us to additionally constrain possible changes in opacity with 
radius. 

If disks are more dusty in their central parts than their outer parts, then 
the disk scale-lengths will increase with inclination.   Indeed, there is 
evidence that disks may in fact be more optically thick at their centres 
(Huizinga \& van Albada 1992; Bosma et al.\ 1992; Jansen et al.\ 1994; 
Giovanelli et al.\ 1994; 
Peletier et al.\ 1995; Beckman et al.\ 1996; Trewhella et al.\ 1997).  
As such galaxies are viewed with increasing inclination, the outer parts of 
the disks will brighten more than the inner parts; this would act to reduce the 
slope of the disk surface brightness profile and therefore increase the disk 
scale-length from its face-on value -- although this effect will be less in the 
near-infrared than the optical. 

On average, in the $K$-band, the decline in opacity with radius must be 
fairly uniform from the 
disk centre to outer boundary.  If this were not the case, the surface brightness 
profiles of the inclined disks should be seen to systematically deviate from an 
exponential light distribution (having S\'{e}rsic shape parameters $n$$>$1) and 
possess kinks and gradient changes due to uneven surface brightness increases.  
Of course some profiles do possess such features, and some display a brightening
of the outer disk profile above that defined by the exponential behaviour of
the inner disk profile.  This could be a signature of a disk which is 
transparent only in the outer parts and brightening with inclination.
However, such galaxies in the sample which behave this way have no preference
for higher inclinations, and the reason for such gradient changes therefore
seems likely to be due to other causes, perhaps resulting from an under-estimation 
of the sky-background level.  The over-whelming majority of $K$-band 
disk profiles in the sample are well described by an exponential law.  

How realistic can changes to the scale-length be for a uniform decrease in
opacity with radius?  Obviously the greatest change will occur if the centre 
of the disk is opaque and the outer, measured parts of the disk are fully 
transparent.  For a galaxy observed at some random inclination, with a central 
disk opacity $C$ and the outer disk fully transparent (i.e.\ $C$$=$1), one can 
compute the change in scale-length, due to inclination, from the face-on value.
For a disk with a measured scale-length $h_{{\rm obs}}$ derived from its light 
profile data which extends to $x$ scale-lengths, the face-on scale-length is 
given by the expression 
\begin{equation}
h_{{\rm face-on}}=h_{{\rm obs}}\frac{x}{x+\Delta\mu/1.086}, \label{eqn2}
\end{equation}
where $\Delta\mu$$=$(1-$C$)$\times$2.5$\log(a/b)$.
For the high-inclination sample of disk galaxies, the average maximum reduction 
to the disk scale-length which can be achieved assuming radial differences in the 
transparency of the disk (with $C$$=$1 at the outer radius, and $C$$=$0 at the 
inner radius) is only 23$\pm$6\%.  This is not enough to bring the low- and 
high-inclination galaxies into agreement in Figure~\ref{fig2}; in fact, it is short
by about a factor of ten.  The disk scale-lengths of the low-inclination galaxies 
undergo an average reduction of 6$\pm$4\% when corrected, under the previous assumption, 
to their face-on value.

Any surface brightness and scale-length inclination correction to the face-on
value will of course affect the high-inclination sample the most.  It is therefore 
of interest to see how the scatter about the $\mu_{0,K}$--$\log h$ relation changes 
in this sample when both of these inclination corrections are applied.  Changing the 
central value of $C$ and correcting $\mu_{0,K}$ and $\log h$ using equations~\ref{eqn1} 
and ~\ref{eqn2} (i.e.\ assuming the outer disk is fully transparent), an ordinary 
least squares bisector line of regression was fitted to the data.  A bisector line 
of regression treats both variables equally, as opposed to the ordinary least squares 
regression of $\mu_{0,K}$ on $\log h$ which was used in the previous Section where the
scatter was considered only to arise from the surface brightness term (see Feigelson 
\& Babu 1994 for a discussion). 
The perpendicular scatter, and also the vertical scatter in $\mu_{0,K}$, were observed
to steadily decrease to a minimum as the central disk value for $C$ approached 1.  
This was also the case when an orthogonal regression line was used.  
When fitting a bisector regression line to the combined low- and high-inclination 
galaxy sample, the minimum perpendicular scatter, and also vertical scatter in 
$\mu_{0,K}$, was achieved when $C$$=$0.9 -- a result independent of the two 
inclination bins. 

To explore this further, a bisector line of regression was individually fitted to 
each of the two data samples and the Working-Hotelling method adapted to compute 
the central value of $C$ which gives the smallest perpendicular offset between 
the two samples in the $\mu_{0,K}$-$\log h$ plane.  
Correcting $\mu_{0,K}$ and changing the scale-length assuming the maximum possible opacity 
gradient, when using the 59 low-inclination galaxies as the calibration set in the 
Working-Hotelling method, the minimum offset occurred when $C$$=$0.89.  Using the 
29 high-inclination galaxies as the calibration set, this occurred when $C$$=$0.97. 

To check for differences arising between the ordinary least squares regression
of $\mu_{0,K}$ on $\log h$ and the bisector method or regression, the 
bisector method of regression was again applied to the data while the 
scale-lengths were held constant, and only corrections to the surface 
brightness applied -- as was done previously.  The minimum perpendicular 
offset between the low- and high-inclination galaxies was found to occur 
when $C$$=$0.91 and 0.98 when the low-inclination and high-inclination 
galaxies were respectively used as the calibration data set.

\section{Discussion}

One of the problems, and hence some of the uncertainty today, with some of the 
past surface brightness inclination tests was that they used the `mean' surface 
brightness ($<$$\mu$$>$) within some fixed isophotal radius.  Consequently, one 
did not know if changes to the mean surface brightness, with inclination, were 
due to changes to the enclosed integrated magnitude or due to changes to the 
isophotal radius (Cho\l oniewski 1991; Burstein, Haynes \& Faber 1991; 
Huizinga \& van Albada 1992).  
With central surface brightnesses ($\mu_{0}$) this is not a problem.  
Furthermore, the selection criteria biases from either a magnitude-limited 
or diameter-limited sample, which have previously been illustrated to operate 
in the apparent magnitude -- apparent diameter ($B_T$--$\log D$) diagram (Burstein 
et al.\ 1991), do not apply to the $\mu_{0}$--$\log D$ or $\mu_{0}$--$\log h$ 
diagram.  For example, 
although the use of apparent magnitudes from a diameter-limited sample 
of galaxies would bias one to conclude that disks are optically thick 
(Burstein et al.\ 1991), this is not the case when using central surface 
brightnesses -- as observed in this investigation.  

Increasing the inclination of optically thin disks from face-on,
their isophotal radii increase while their magnitudes remain constant; if
they are optically thick, their magnitudes decrease while their isophotal
radii remain constant.  Independent of the opacity of the disk, this
situation results in `inclined' galaxies changing positions in the
$B_T$--$\log D$ plane from their face-on values and leaving one uncertain
as to whether the apparent magnitude,
the isophotal diameter or a combination of both has changed.  Cho\l oniewski (1991)
and Burstein et al.\ (1991) tried to resolve this dilemma by using distance
information, but Davies et al.\ (1995) showed the presence of the
redshift-cutoff selection criteria left this mean surface brightness inclination
test unable to tell one anything about the transparency of disks.  This dilemma,
however, does not exist in the $\mu_{0}$--$\log D$ diagram. 
If disks are optically thick, then neither $\mu_{0}$ or
$\log D$ will change with inclination.  If disks are optically thin then they will
move towards brighter $\mu_{0}$ and larger $\log D$ values as they are viewed at
greater inclination angles.  Therefore, any positional
correlation with inclination in the $\mu_{0}$--$\log D$ diagram will indicate
that disks are at least semi-transparent; a lack of any such trend will indicate
they are opaque.  The strength of any such trend depending on the inclination
of the sample (which is known) and the apparent opacity of the disk (which one
could then measure).
The situation is even simpler here because scale-lengths, rather than
isophotal diameters, have been used, and for a globally constant disk
opacity the former will not change with inclination - although possible
radial changes in the opacity of the disk have been explored here.

For transparent galaxies, as the inclination angle increases the 
isophotal diameters will increase.  Consequently, the high-inclination
galaxy sample may contain galaxies which are intrinsically smaller
than the galaxies in the low-inclination sample.  Although, as noted by
Huizinga (1994), this may not be such a problem for galaxies selected from the UGC.
Their diameters were estimated by eye to the faintest structures visible 
on the photographic plates.  Given that these structures are probably 
H$_{\rm II}$ regions whose surface brightness does not change with inclination, 
the diameter inclination dependence is probably small.  Anyway, 
plotting central surface brightness against disk scale-length should account 
for any selection bias due to galaxy size in this inclination test.  

This investigation is free from criticisms levelled at previous studies 
which used central disk surface brightnesses without accurately considering 
the contribution of the bulge light to the disk profile.
Kormendy (1977) and Davies (1990) have shown how the linear part of the 
surface brightness profile can be contaminated by bulge light, and how fitting
to this part of the profile alone 
can result in erroneous estimates of the central disk surface brightness.  
In the analysis presented here, 
S\'{e}rsic $r^{1/n}$ models convolved with a Gaussian PSF have been 
fitted to the bulge while an exponential model, also 
convolved with the PSF, was simultaneously fitted to the disk.  
For the low-inclination sample, the FWHM of the PSF had been individually obtained
for each galaxy from stars on the image frames (de Jong 1996a).  For the 
high-inclination sample, a value of 1.0 arcsec was used -- the actual seeing 
varied from 0.8$\arcsec$ to 1.2$\arcsec$  but this had little affect on the results 
of the disk parameters (see also Khosroshahi et al.\ 2000). 

There have been a few near-infrared investigations, employing different techniques, 
to measure the value of $C$ required
for the inclination correction.  Using 1.65$\mu$m ($H$-band) aperture
photometry, Peletier \& Willner (1992) and Boselli \& Gavazzi (1994) presented 
evidence that disks behaved in a semi-transparent way: with 
$C_H$$\geq$0.65 and $C_H$$=$0.5-0.9 respectively. 
A more recent study using a two-dimensional  bulge/disk decomposition 
found $C_H$$=$0.60$\pm$0.14 (Moriondo, Giovanelli \& Haynes 1998). 
With a $K'$ study, Tully \& Verheijen (1997) found that the minimum scatter in 
the central disk surface brightness for their high surface brightness (HSB) 
galaxies occurred when $C_{K'}$$=$1, in agreement with this papers $K$-band analysis. 
Given that near-infrared (2.2-${\rm\mu}$m)  light passes largely undisturbed 
through dust clouds, the line-of-sight depth into (and through?) a spiral 
galaxy disk is expected to increase as one views spiral galaxies, and hence 
disks, at greater inclinations. 
It is therefore reassuring to find that the observed central disk surface 
brightness does increase with galaxy inclination in the $K$-band.

The decision to use spiral galaxies with morphological types $T$$\leq$5 was 
determined by two factors.  Firstly, the high-inclination galaxy sample were all
of this type.  It should perhaps be mentioned that difficulties in matching 
morphological type at different inclinations may introduce some small bias, but
given the inherent uncertainty in this business (Lahav et al.\ 1995) this is not
expected to be important for this sample.  
The other reason for using the early-type galaxies is that the 
late-type spirals do not follow the $\mu_{0,K}$--$\log h$ relation evident for
the early-type disk galaxies.  
For a given scale-length, galaxy types later than Sc can be considerably fainter
than the earlier type spirals.
To show this, Figure~\ref{fig2} displays the data 
for the low-inclination disk galaxies having type $T$$>$5.
This data has not been included in any of the calculations presented previously. 

Further evidence to segregate the late-type and early-type disk galaxies comes from 
work by Tully \& Verheijen (1997) who show that the S0-Sc and Scd-Sm-Irr galaxies 
have markedly different central surface brightness values, not just as a function 
of Hubble type but also in the $\mu_{0}$--$\log h$ plane (but see Bell \& de 
Blok 2000).  Considering Figure~\ref{fig2} from this paper with Figure 2 from 
van den Bosch (1998) -- which shows a $V$-band $\mu_{0}$--$\log h$ relationship 
for all types of galaxy disks -- one can see that the late-type spirals from 
the sample of de Jong \& van der Kruit reside in the same part of the diagram 
as the low surface brightness galaxies.  Previous inclination tests which have 
looked for a correlation 
between $\mu_{0}$ and $\log(b/a)$ may therefore suffer confusion/noise because of 
inclusion of late-type galaxies ($T$$>$5) and because of the relationship 
between surface brightness and scale-length amongst the early-type spirals.  

Could selection effects be responsible for this trend between $\mu_{0,K}$ and $\log h$
by preventing early-type galaxies from occupying the upper-right or lower-left of
the $\mu_{0,K}$--$\log h$ diagram?
The selection criteria are such that they favour the detection of galaxies with
large scale-lengths and luminous central surface brightnesses.  At a fixed scale-length,
the probability of detecting galaxies that have central surface brightnesses which are
brighter than those observed along the $\mu_{0,K}$--$\log h$ relation is actually greater
than the probability of detecting those galaxies at the same scale-length which
define the observed relation.  Therefore, the upper envelope in the $\mu_{0,K}$--$\log h$
diagram reflects a true drop-off in the number density of
bigger, brighter spiral galaxies, and has been noted before.

\begin{figure}
\centerline{\psfig{figure=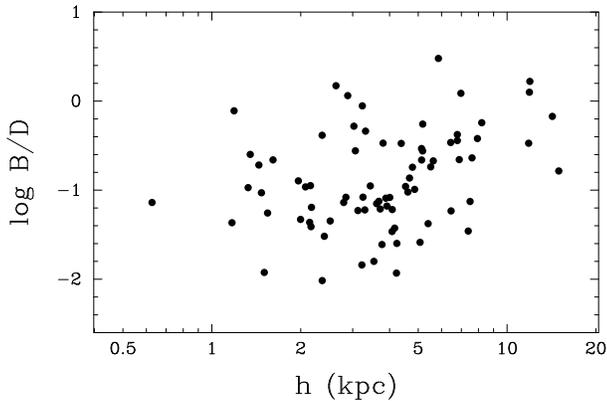,width=8.0cm,angle=-90}}
\caption{Bulge-to-disk ratio versus disk scale-length for galaxies with 
morphological type $\leq$Sc.  The full $C$$=$1 inclination correction has 
been applied to the disk (see text for details).  A Hubble constant of 75 
km s$^{-1}$ Mpc$^{-1}$ has been used.}
\label{fig3}
\end{figure}

Grosb\o l (1985) suggested that it might be due to an upper constant
luminosity limit, in which case the slope should be 5.
However, his photographic data appears more consistent with a slope of $\sim$3.
Figure~\ref{fig3} shows that for the early-type spiral
galaxies, the $B/D$ luminosity ratio is, if anything, smaller for the smaller 
scale-length galaxies.  Therefore, the $\mu_{0,K}$--$\log h$ relation clearly 
does not delineate a constant upper luminosity boundary to the luminosity 
function -- even when the presence of the bulge light is factored in.  
Fitting a line of constant luminosity in the $\mu_{0}$--$\log h$ diagram with
a slope of 2.5 instead of 5, Kent (1985) mistakingly attributed the
$\mu_{0}$--$\log h$ correlation to a limited range of disk luminosities in his sample,
Using the $I$-band Sb-Sd galaxy data from Byun's (1992) thesis, the slope of this
envelope is $\sim$1.9 (see also Reshetnikov 2000), and a somewhat smaller value of 1.5
is obtained from the composite data of Dalcanton, Spergel \& Summers (1997, their
figure 4). 
Evstigneeva \& Reshetnikov (1999) have claimed, from a compilation of 356 galaxies, 
that normal high surface brightness spiral disks follow a trend such that 
$\mu_0\propto 2.5\log h$ with a correlation coefficient of 0.45 and a scatter of 
$\sim$1 mag arcsec$^{-2}$.  After submission, the author discovered that 
M\"ollenhoff \& Heidt (2001) report a slope of 2.02$\pm$0.64 ($r=0.61$) in 
the near-infrared for a sample of 40 bright, low-inclination, early-type 
($leq$Sc) spiral galaxies without bars, in complete agreement with the 
results obtained here (see also Seigar \& James 1998; Moriondo, Giovanardi, 
\& Hunt 1998 and Khosroshahi et al.\ 2000).  

Recent studies of bulge dominated low surface brightness (LSB) galaxies 
(Beijersbergen, de Blok \& van der Hulst 1999) have shown that they
appear to continue the trend defined by the normal early-type high 
surface brightness (HSB) galaxies which have been used here.
With considerably more scatter still, a similar trend exists not just for 
the ordinary high surface brightness spiral galaxy disks, but also for 
the low surface brightness disks, disky ellipticals and HST resolved nuclear 
disks (Scorza \& van den Bosch 1998; Reshetnikov 2000).  
For a constant velocity, the virial theorem, which can be expressed
as $v^2\propto Ih$, predicts a slope of 2.5 in the
$\mu_{0}$--$\log h$ diagram.  How, and whether or not, this could be the 
limiting criteria is not clear.

In passing, it is noted that the presence of the high-inclination galaxies 
located 1 mag above the upper envelope to the $\mu_{0,K}$--$\log h$ relation 
defined by the low-inclination sample is strong evidence for the need of the 
inclination correction.

The occupation of the late-type spirals in the lower-left of the 
$\mu_{0,K}$--$\log h$ diagram, in the region of lower rotational velocity, 
is in accord with observational results (Zaritsky, 1993, his figure 2).
The selection criteria do however somewhat restrict the galaxy sample from 
occupying this part of the diagram. 
The dotted lines in the upper panel of Figure~\ref{fig2}
reveal the nature of the selection criteria and show the completeness of the
sample to the indicated distances (15, 40, 100 Mpc; $H_{\rm 0}$$=$75 km s$^{-1}$
Mpc$^{-1}$).  Assuming perfectly exponential
disks, these lines show the 2.0$\arcmin$ diameter limit at the 22.45 $K$-mag
arcsec$^{-2}$ isophote.  (The average surface brightness where the UGC red diameters
were measured is 24.7 $R$-mag arcsec$^{-2}$, and the mean $R$-$K$ colour at
the 60$\arcsec$ radius for those galaxies in the low-inclination sample with
photometric data and morphological type 1$\leq$$T$$\leq$5 is 2.25$\pm$0.07.)
Galaxies near the 100 Mpc line had (100/15)$^3$$\sim$300 times more chance of being
included in the sample than those near the 15 Mpc line.
For any ($\mu_{0,K}$,$\log h$) parameter combination between two close pairs of these
curved lines, there is an equal
probability of observing a galaxy if it exists in, at least, the local Universe.
Therefore, within any such curved region, the absence of early-type disk galaxies below
the $\mu_{0,K}$--$\log h$ relation indicates that they do not exist here in as
great a number density as the population on the observed $\mu_{0,K}$--$\log h$
relation and within the same curved bands.
The presence of the late-type galaxy population in the lower part of the
$\mu_{0,K}$--$\log h$ diagram suggests that these galaxies exist with a greater
number density than the early-type spirals in this domain of the parameter 
space.\footnote{The late-type galaxies are some 0.30$\pm$0.16 mag bluer in 
$R$-$K$ than the early-type galaxies and so the selection criteria lines shown
in Figure~\ref{fig2} for the early-type spirals will be different by this amount
for the late-type spirals.}  
The inclusion of disk-dominated low surface brightness galaxies
is known, at least at shorter wavelengths, to fill in the lower left of the 
diagram (Bothun et al.\ 1997, their figure 4; Dalcanton et al.\ 1997; McGaugh 1998). 
Whether or not, in the $K$-band, early-type spiral galaxies reside in the lower
left of the $\mu_{0}$--$\log h$ diagram in any great number density cannot be 
answered with the present data set.  Consequently, while the observed 
$\mu_{0}$--$\log h$ relation may reflect the location to the bulk of the 
early-type galaxy population, it may also only reflect the upper 
envelope -- which is, in itself, still of interest.  
It would be of benefit to extend the current data sample to smaller disks
and fainter surface brightness levels. 

\section{Main conclusions}

The well known modification to Freeman's law to incorporate fainter 
central disk surface brightnesses is, in some respect, partially quantified.
A relation is presented for the disks of early-type spiral galaxies.  
Their central disk surface brightnesses, at least
in the $K$-band, are shown to increase (that is, become fainter) 
with increasing disk scale-length, such that $\mu_{0,K}\propto 2\log h$. 
This relation is not adhered to by the late-type ($geq$Scd) spiral galaxies.

It is found that the disks of early-type spiral galaxies ($leq$Sc) should 
be considered transparent for the sake of inclination corrections to their 
observed photometric parameters.   From a number of tests, 
the optimal value of $C$ in the standard inclination correction formula 
$2.5C\log(b/a)$, where $b/a$ is the minor-to-major axis ratio, resides 
between 0.9 and 1.0.  This implies that the maximum correction, in the 
confines of the inclination correction formula, 
should be applied to the observed $K$-band disk surface brightness, 
and simultaneously, no corrective term for dust should be applied to 
the observed K-band magnitudes.

\section{acknowledgements}

I am grateful to John Beckman and Marc Balcells for their reading and comments 
on this paper.  I also wish to thank Peter Erwin for enjoyable discussions, 
Eric Feigelson for making the computer code SLOPES and CALIB publicly available
and Ren\'{e}e Kraan-Korteweg for sending me her Virgo-centric inflow code. 
I thank Gregory Bothun and R.\ de Jong for their input and suggestions,
which helped to improve this paper. 
This research has made use of the NASA/IPAC Extragalactic Database (NED)
which is operated by the Jet Propulsion Laboratory, California Institute
of Technology, under contract with the National Aeronautics and Space
Administration.

\label{lastpage}

\end{document}